\documentclass[a4paper,11pt]{article}
\usepackage[utf8]{inputenc}

\usepackage{bm}
\usepackage{comment} 

\usepackage[top=15truemm,bottom=20truemm,left=20truemm,right=20truemm]{geometry}
\usepackage[colorlinks=true,urlcolor=blue,anchorcolor=black,citecolor=blue,linkcolor=black,filecolor=black,menucolor=black,pagecolor=black,linktocpage=true,pdfproducer=medialab,pdfa=true]{hyperref}
\usepackage{graphicx}
\usepackage{amsmath,latexsym,amssymb,mathrsfs,ascmac,physics,mathtools}
\numberwithin{equation}{section}
\usepackage[affil-it]{authblk}
\usepackage{cite}
\usepackage{lineno}

\begin{document}

\title{Higher curvature corrections to the black hole Wheeler-DeWitt equation and the annihilation to nothing scenario}

\author{Takamasa Kanai\thanks{kanai@kochi-ct.ac.jp}}

\affil{Department of Social Design Engineering,
National Institute of Technology (KOSEN), Kochi College,
200-1 Monobe Otsu, Nankoku, Kochi, 783-8508, Japan}

\date{}

\maketitle

\begin{abstract}
We revisit Yeom’s annihilation-to-nothing scenario using a modified Wheeler–DeWitt (WDW) equation incorporating higher-curvature corrections. We show that, once these corrections are taken into account, the WDW wave function exhibits severe divergences arising from contributions near the classical singularity. These divergences indicate that the low-energy effective field theory (EFT) description breaks down in this regime.

Given that general relativity (GR) itself is merely a low-energy effective field theory (EFT) of an underlying ultraviolet (UV) theory, our results suggest that any attempted resolution of the black hole singularity cannot be reliably discussed within the EFT framework. Our analysis does not contradict Yeom’s conjecture, but emphasizes that the annihilation-to-nothing scenario should be discussed within a UV-complete theoretical framework. It further clarifies that any genuine resolution of the singularity necessarily requires a framework capable of appropriately describing ultraviolet physics, such as degrees of freedom beyond those captured by GR or dynamics consistently defined up to arbitrarily high energy scales.
\end{abstract}

\section{Introduction}

The resolution of black hole singularities remains one of the central open problems in theoretical physics. General relativity (GR), despite its remarkable success as a classical theory of gravitation, inevitably predicts the formation of curvature singularities inside black holes \cite{Penrose:1964wq,Hawking:1970zqf,Hawking:1973uf}. Since such singularities signal the breakdown of the classical description, it is widely believed that a quantum theory of gravity is required to address this issue. 

One of the well-studied approaches to the singularity problem is quantum cosmology based on the Wheeler-DeWitt (WDW) equation \cite{Halliwell:1989myn,Kiefer:2013jqa,Kiefer:2008sw,Kiefer:2025udf}. Inside the black hole horizon, where the roles of the time and radial coordinates are interchanged, the geometry can be described by a minisuperspace model analogous to a homogeneous cosmology. The WDW equation is particularly suitable for this setting, as it naturally accommodates the time–radial coordinate interchange and allows a canonical analysis of quantum gravitational effects in the black hole interior \cite{Bouhmadi-Lopez:2019kkt,Yeom:2019csm,Yeom:2021bpd,Brahma:2021xjy,Perry:2021mch,Perry:2021udd,Hartnoll:2022snh}. Within this framework, Yeom and collaborators proposed the so-called \emph{Annihilation-to-nothing} scenario~\cite{Bouhmadi-Lopez:2019kkt,Yeom:2019csm,Yeom:2021bpd,Brahma:2021xjy}, in which two classical branches of spacetime---one evolving from the horizon and the other from the singularity---annihilate at a hypersurface inside the horizon. This interpretation provides a possible mechanism for the resolution of black hole singularities within the WDW formalism. However, whether such a mechanism survives once more fundamental quantum effects are included remains an open question.

In this broader context, nonperturbative approaches, such as loop quantum gravity \cite{Modesto:2004wm,Ashtekar:2005qt,Corichi:2015xia,Sartini:2020ycs,Geiller:2020xze,Ongole:2022rqi} or string theoretic constructions \cite{Polchinski:1998rq,Polchinski:1998rr}, suggest that singularity resolution can occur due to fundamentally quantum effects. However, it is unclear whether such effects can be captured within a low energy effective field theory (EFT) framework. From the EFT perspective \cite{Weinberg:1978kz,Donoghue:1993eb,Donoghue:1994dn,Burgess:2003jk}, GR is viewed as the low energy limit of a more fundamental ultraviolet (UV) theory. Consequently, it is unnatural to expect that singularity resolution can be achieved solely within the canonical quantization of the GR framework, as its description inevitably breaks down near the Planck scale, where higher curvature operators and new physical degrees of freedom become significant.

In this paper, we revisit Yeom’s \emph{annihilation-to-nothing} interpretation by introducing higher-curvature corrections within the effective field theory (EFT) framework, treating general relativity (GR) as an effective field theory valid below a ultraviolet (UV) cutoff. By analyzing the modified Wheeler–DeWitt (WDW) equation derived from this extended action, we re-examine Yeom’s wave packet solutions and their behavior in the presence of higher-curvature terms \cite{Endlich:2017tqa}.

We show that the inclusion of EFT corrections generically leads to severe divergences arising from contributions near the classical singularity. These divergences signal a breakdown of the perturbative EFT expansion in this regime, and the divergence itself constitutes a strong counterargument to the robustness of the annihilation-to-nothing scenario within the EFT framework. Our results demonstrate that the annihilation-to-nothing scenario should be discussed within a UV-complete theoretical framework. By a UV-complete theory of quantum gravity, we mean a framework in which the breakdown of the low-energy effective field theory near the singularity is not addressed by a finite-order truncation or a modification of the effective description, but is instead overcome through the emergence of genuinely new degrees of freedom or through dynamics that are consistently defined up to arbitrarily high energy scales. The explicit breakdown of the EFT expansion at the singular point thus supports our central conclusion that the physics of the black hole singularity lies beyond the domain of validity of low-energy EFT.

There have been several attempts to address singularity resolution within effective frameworks \cite{Bosso:2019ljf,Bosso:2020ztk,Bosso:2023fnb,Melchor:2023rqd}. These studies demonstrate that specific modifications to the Einstein equations or the WDW equation can lead to non-singular bounce solutions. However, the effective dynamics considered in those motivated works by loop quantum gravity \cite{Modesto:2004wm,Ashtekar:2005qt,Corichi:2015xia,Sartini:2020ycs,Geiller:2020xze,Ongole:2022rqi} and string theory \cite{Polchinski:1998rq,Polchinski:1998rr} go beyond a simple perturbative curvature expansion and incorporate certain nonperturbative features. Nevertheless, it remains an effective description whose domain of validity near the Planck-scale curvature is not fully under theoretical control. In the vicinity of the classical singularity, different higher-curvature or nonlocal corrections may become relevant, and conclusions drawn from a specific effective model may therefore lack robustness. Therefore, we argue that a genuine resolution of the singularity requires a UV-complete theory of quantum gravity to determine the full non-perturbative behavior. 

The novelty of this work lies in providing the first systematic EFT extension of the Wheeler–DeWitt equation for black hole interiors. By consistently incorporating higher-curvature corrections, we show that Yeom’s annihilation-to-nothing scenario fails to hold within the effective field theory framework. This suggests that the apparent singularity resolution suggested by the classical WDW analysis cannot be realized within a low-energy effective description, and instead requires a framework capable of appropriately describing ultraviolet physics, such as degrees of freedom beyond those captured by GR or a framework whose dynamics remain well-defined at arbitrarily high energies.

The organization of this paper is as follows. In Sec.~2, we briefly review the WDW formulation inside black hole horizons and Yeom’s \emph{Annihilation-to-nothing} interpretation. In Sec.~3, we discuss Yeom’s proposal by introducing higher-derivative corrections at the classical level within the EFT framework. In Sec.~4, we analyze the wave functions by incorporating the higher-derivative terms as quantum corrections in EFT and discuss their implications for Yeom’s interpretation. Finally, Sec.~5 presents our conclusion and outlook for future research.

Unless specified otherwise, we use the natural unit system, where the speed of light $c=1$ and the Dirac constant $\hbar=1$.

\section{Review of Wheeler-DeWitt wave function inside black holes and Yeom’s interpretation }

We review the derivation of the minisuperspace Wheeler DeWitt (WDW) equation inside the horizon of spherically symmetric black holes in arbitrary dimensions. According to Birkhoff’s theorem \cite{Misner:1973prb}, any spherically symmetric vacuum solution of the Einstein equation must be static and is uniquely given by the Schwarzschild--Tangherlini metric in $D$ dimensions. Therefore, the interior region of a Schwarzschild black hole provides a representative setup for studying the minisuperspace dynamics. We further review Yeom's interpretation of the WDW wave function for black hole interior in this section.

\subsection{Wheeler-DeWitt equation inside a spherically symmetric black hole}

We start from the Einstein-Hilbert action in $D$-dimensional spacetime,
\begin{equation}
S = \frac{1}{16\pi G}\int_M d^Dx \sqrt{-g} R,
\end{equation}
where $g$ is the determinant of the metric $g_{ab}$, $R$ is the $D$-dimensional Ricci scalar. 

Inside the horizon, the Kantowski–Sachs–type metric is adopted\cite{Kantowski:1966te},
\begin{equation}
ds^2 = - e^{2n(t)} dt^2 + e^{2\alpha(t)} dr^2 + r_s^2 e^{2\beta(t)} d\Omega_{D-2}^2 ,
\end{equation}
where $d\Omega_{D-2}^2$ is the line element of the $(D-2)$-sphere with unit radius and $r_s$ a constant of dimension length.  
This interior metric is related to the Schwarzschild–Tangherlini metric by
\begin{equation}
e^{2\beta} = \frac{t^2}{r_s^2}, \qquad
e^{2\alpha} = \frac{r_s^{D-3}}{t^{D-3}} - 1, \qquad
e^{2n} = \!\left(\frac{r_s^{D-3}}{t^{D-3}} - 1\right)^{-1},
\end{equation}
where time and radial coordinates are exchanged relative to the exterior region.

Choosing the lapse function as
\begin{equation}
n(t) = \alpha(t) + (D-2)\beta(t),
\end{equation}
which can be achieved by a gauge transformation, and substituting the metric into the action, the minisuperspace reduction yields
\begin{equation}
S \propto \!\int dt
  \!\left[
  -2\dot{\alpha}\dot{\beta}-(D-3)\dot{\beta}^2
  + (D-3) r_s^{2(D-3)} e^{2\alpha+2(D-3)\beta}
  \right].
\end{equation}
The action can further be rewritten in terms of the variables $X=\alpha$ and $Y=\alpha+(D-3)\beta$ as
\begin{equation}
S \propto \!\int dt
  \!\left[
  \frac{1}{D-3}(\dot{X}^2 - \dot{Y}^2)
  + (D-3) r_s^{2(D-3)} e^{2Y}
  \right].
\end{equation}
Therefore, we obtain the canonical Hamiltonian
\begin{equation}
H = (D-3)\left[\frac{1}{4}(\Pi_X^2 - \Pi_Y^2)- r_s^{2(D-3)} e^{2Y}\right],
\end{equation}
where $\Pi_X$ and $\Pi_Y$ are the canonical conjugate momenta to $X$ and $Y$, respectively.
Through canonical quantization, the canonical momenta are replaced by differential operators acting on the wave function,
\begin{equation}
\Pi_X \rightarrow \hat{\Pi}_X = -i\frac{\partial}{\partial X}, \qquad
\Pi_Y \rightarrow \hat{\Pi}_Y = -i\frac{\partial}{\partial Y},
\end{equation}
resulting in the Wheeler–DeWitt (WDW) equation in the minisuperspace \cite{Halliwell:1989myn,Kiefer:2013jqa,Kiefer:2008sw,Kiefer:2025udf}.
The WDW equation arises from the Hamiltonian constraint $\hat{\mathcal{H}}\Psi = 0$, which requires that the wave function of the universe satisfy this constraint in quantum gravity.
Thus, the minisuperspace WDW equation can be written as
\begin{equation}
\label{eq:WDW-spherical}
\left(\frac{\partial^2}{\partial X^2}- \frac{\partial^2}{\partial Y^2}+ 4 r_s^{2(D-3)} e^{2Y}\right)\Psi(X,Y)=0 .
\end{equation}
This form is dimension-independent up to the power of $r_s$.

In this coordinate system, the region $X,Y\rightarrow-\infty$ represents the event horizon, whereas the limit $X\rightarrow\infty$ and $Y\rightarrow-\infty$ corresponds to the spacetime singularity.

\subsection{Solutions and Yeom's interpretation}

The fundamental solution of (\ref{eq:WDW-spherical}) is
\begin{equation}
\psi_k(X,Y) = e^{-i k X} K_{i k}(2r_s e^{Y}),
\end{equation}
and general solutions are given by
\begin{equation}
\Psi(X,Y) = \int_{-\infty}^{\infty} f(k)\,\psi_k(X,Y)\, dk,
\end{equation}
where $f(k)$ is an amplitude function.

Using the asymptotic form of the modified Bessel function, one can construct Gaussian or analytical wave packets. For example, choosing $f(k)=i k$ yields
\begin{equation}
\Psi_1(X,Y)
= 2\pi r_s^{D-3} e^{Y} \sinh X
  \, e^{-2r_s^{D-3} e^{Y}\cosh X}.
\end{equation}

\begin{figure}[htbp]
   \centering
   \includegraphics[width=0.7\linewidth]{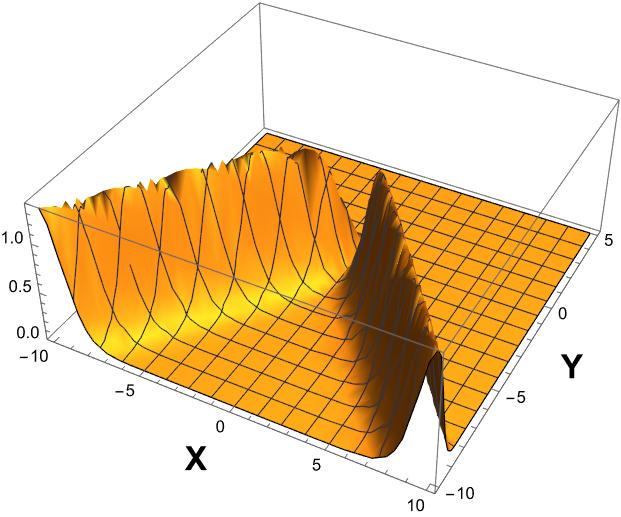}
   \caption{Probability density of the wave function for $r_s=1$. Adapted from Ref.~\cite{Bouhmadi-Lopez:2019kkt}.}
   \label{fig1}
\end{figure}

The probability density $\rho=|\Psi_1|^2$ shows a peak following the classical trajectory
\begin{equation}
e^{Y}\cosh X = \mathrm{const.},
\end{equation}
and it vanishes near $X=0$, where neither the horizon nor the singularity exists. This behavior is interpreted as the ``Annihilation-to-nothing'' process
proposed by Yeom \emph{et al.} \cite{Bouhmadi-Lopez:2019kkt,Yeom:2019csm,Yeom:2021bpd,Brahma:2021xjy}, representing the annihilation of two classical spacetime branches with opposite time arrows inside the horizon. The behavior of the probability density on $X-Y$ plane is shown in Fig. \ref{fig1}.

To summarize this section, the WDW wave packet inside spherical black holes exhibits behavior consistent with an annihilation-to-nothing process, thereby supporting Yeom’s interpretation. In this interpretation, wave packets propagating along opposite arrows of time are assumed to collide in the vicinity of the singularity and undergo a wave packet annihilation process. As a result, the WDW wave function describing the black hole interior vanishes, and the spacetime is interpreted as being connected to another spacetime region free of singularities. Therefore, the resolution of the singularity is understood as a quantum phenomenon realized through the interference of the wave function of the universe. Related analyses have also been carried out for topological black holes using the scalar-type WDW equation and for black holes described by the Dirac-type WDW equation \cite{Kan:2021yoh,Kan:2021fmw,Kan:2022ism}. For topological black holes, the results remain compatible with Yeom’s interpretation, whereas the Dirac-type formulation yields a novel variant of the annihilation picture.

General relativity is commonly understood as a low-energy EFT, and its applicability to the problem of singularity resolution is therefore intrinsically limited. From this perspective, the apparent singularity resolution observed in the classical Wheeler–DeWitt analysis should be interpreted with caution. Indeed, as shown by the results of this paper, once higher-derivative corrections that generically arise in the effective field theory of gravity are taken into account, divergent contributions appear, which suggests a breakdown of the EFT description itself.

In the following, we examine this issue explicitly. In Sec.~3, we introduce classical higher-derivative corrections, while in Sec.~4 we analyze quantum effects arising from curvature-squared terms that vanish at the classical level. Through these analyses, we reassess Yeom’s annihilation-to-nothing interpretation within the EFT framework.

\section{Wheeler-DeWitt wave function with classical higher curvature corrections}

In this section, we consider classical corrections to the WDW wave function in the black-hole interior within the framework of effective field theory. Since we focus on vacuum solutions, the Ricci tensor vanishes identically, and the action can therefore be constructed solely from the Riemann tensor. Modulo tensor identities~\cite{Fulling:1992vm}, the only independent fourth-derivative invariant that does not involve the Ricci tensor, is $R_{\mu\nu\rho\sigma}R^{\mu\nu\rho\sigma}$; this is field-redefinition equivalent to the Gauss–Bonnet combination $R^2-4R_{\mu\nu}R^{\mu\nu}+R_{\mu\nu\rho\sigma}R^{\mu\nu\rho\sigma}$, which in four dimensions is topological and therefore does not affect the bulk equations of motion. At the next order, the independent Riemann-constructed operators excluding topological terms are $\nabla_{\alpha}R_{\mu\nu\rho\sigma}\nabla^{\alpha}R^{\mu\nu\rho\sigma}$ and $R^{\rho\sigma}_{\ \ \mu\nu}R^{\alpha\beta}_{\ \ \rho\sigma}R^{\mu\nu}_{\ \ \alpha\beta}$; the former can be rearranged, up to a total derivative, into a Riemann-cubed term plus Ricci-dependent operators and so can be dropped without loss of generality. Likewise, there are two independent components of the fourth power of Riemann tensors. 

The effective action considered here is not chosen ad hoc. Rather, they are dictated by the general principles of effective field theory: we include all independent higher-derivative operators constructed from the Riemann tensor that are compatible with diffeomorphism invariance, organized according to the number of derivatives. Moreover, these operators are understood as part of a derivative expansion: higher-curvature terms involving more derivatives are assumed to be increasingly suppressed by the cutoff scale. In particular, in four-dimensional vacuum spacetimes the leading nontrivial classical higher-curvature contributions arise at cubic order in the curvature. In Ref.~\cite{Camanho:2014apa}, it was argued that curvature-cubic terms do not necessarily constitute the leading corrections to general relativity. Motivated by this observation, in the present work we include quartic curvature invariants on the same footing and analyze their effects alongside the cubic terms. Thus, we consider the following action in four dimensional spacetime,
\begin{equation}
S=\frac{1}{16\pi G}\int d^4x\sqrt{-g}(R+\gamma R^{\rho\sigma}_{\ \ \mu\nu}R^{\alpha\beta}_{\ \ \rho\sigma}R^{\mu\nu}_{\ \ \alpha\beta}+\eta \mathcal{C}^2+\lambda \tilde{\mathcal{C}}^2),
\end{equation}
where $\mathcal{C}$ and $\tilde{\mathcal{C}}$ are defined by
\begin{eqnarray}
\mathcal{C}&=R_{\mu\nu\rho\sigma}R^{\mu\nu\rho\sigma},\\
\tilde{\mathcal{C}}&=\tilde{R}_{\mu\nu\rho\sigma}R^{\mu\nu\rho\sigma},
\end{eqnarray}
with the dual tensor $\tilde{R}_{\mu\nu\rho\sigma}=\epsilon_{\mu\nu}^{\ \ \alpha\beta}R_{\alpha\beta\rho\sigma}$.

As discussed in the previous section, the metric inside a black hole can be written in the Kantowski–Sachs form,
\begin{equation}
ds^2=g_{\mu\nu}dx^{\mu}dx^{\nu}=-e^{2\alpha(t)+4\beta(t)}dt^2+e^{2\alpha(t)}dr^2+r_s^2e^{2\beta(t)}d\Omega_2^2.
\end{equation}
For the Schwarzschild spacetime, the functions take the specific form
\begin{equation}
e^{2\alpha(t)}=\frac{r_s}{\tau(t)}-1,\ \ \ e^{2\beta(t)}=\frac{\tau^2(t)}{r_s^2}.
\end{equation}
Note that $\tau$ denotes the radial coordinate of the familiar Schwarzschild metric, and in the present case it is taken to be a function of the spacetime coordinate $t$. Then, the higher derivative corrections are
\begin{equation}
R^{\rho\sigma}_{\ \ \mu\nu}R^{\alpha\beta}_{\ \ \rho\sigma}R^{\mu\nu}_{\ \ \alpha\beta}=\frac{12r_s^3}{\tau^9(t)},\ \ \ \mathcal{C}^2=\frac{144r_s^4}{\tau^{12}(t)},\ \ \ \tilde{\mathcal{C}}^2=0.
\end{equation}
Using the variables $X$ and $Y$ introduced in the previous section, these are replaced with the following operators in the quantum Hamiltonian constraint, $\hat{\mathcal{H}}\Psi=0$.
\begin{equation}
R^{\rho\sigma}_{\ \ \mu\nu}R^{\alpha\beta}_{\ \ \rho\sigma}R^{\mu\nu}_{\ \ \alpha\beta}\rightarrow\frac{12}{r_s^6}e^{9X-9Y},\ \ \ \mathcal{C}^2\rightarrow\frac{144}{r_s^8}e^{12X-12Y},\ \ \ \tilde{\mathcal{C}}^2\rightarrow0.
\end{equation}
The WDW equation then takes the following form.
\begin{eqnarray}
&&\hat{\mathcal{H}}\Psi=-\left(\frac{1}{4}\frac{\partial^2}{\partial X^2}-\frac{1}{4}\frac{\partial^2}{\partial Y^2}+r_s^2e^{2\hat{Y}}-\frac{12\gamma}{r_s^4}e^{7X-5Y}-\frac{144\eta}{r_s^6}e^{10X-8Y}\right)\Psi(X,Y)\nonumber\\
&&\qquad=0.
\end{eqnarray}
Expanding the WDW equation perturbatively in terms of the higher curvature couplings $\gamma$ and $\eta$, we separate the zeroth- and first-order contributions as follows:

\begin{eqnarray}
&&\left(\frac{1}{4}\frac{\partial^2}{\partial X^2}-\frac{1}{4}\frac{\partial^2}{\partial Y^2}+r_s^2e^{2\hat{Y}}\right)\Psi^{(0)}(X,Y)=0,\\
\label{eq1}
&&\left(\frac{1}{4}\frac{\partial^2}{\partial X^2}-\frac{1}{4}\frac{\partial^2}{\partial Y^2}+r_s^2e^{2\hat{Y}}\right)\Psi^{(1)}(X,Y)=\left(\frac{12\gamma}{r_s^4}e^{7X-5Y}+\frac{144\eta}{r_s^6}e^{10X-8Y}\right)\Psi^{(0)}(X,Y).
\end{eqnarray}
To facilitate the analysis, we introduce a change of variables $(X,Y)\rightarrow(Z,W)$ such that the WDW operator, including both the kinetic and potential terms, is brought to a standard form up to an overall conformal factor. In terms of these new variables, the operator takes the form
\begin{eqnarray}
\left(\frac{1}{4}\frac{\partial^2}{\partial X^2}-\frac{1}{4}\frac{\partial^2}{\partial Y^2}+r_s^2e^{2\hat{Y}}\right)=(W^2-Z^2)\left(\frac{1}{4}\frac{\partial^2}{\partial Z^2}-\frac{1}{4}\frac{\partial^2}{\partial W^2}-1\right),
\end{eqnarray}
where the new coordinates are defined by
\begin{equation}
\label{coordinate trans.}
Z=r_se^Y\cosh X,\ \ \ \ W=r_se^Y\sinh X,
\end{equation}
which can be inverted as
\begin{equation}
e^Y=\frac{\sqrt{Z^2-W^2}}{r_s},\ \ \ e^X=\frac{Z+W}{\sqrt{Z^2-W^2}}.
\end{equation}
This transformation maps the minisuperspace into the wedge defined by $Z>|W|$.

Substituting these relations into Eq.~(\ref{eq1}), the first-order wave function $\Psi^{(1)}$ can be expressed in terms of the Green’s function $G(Z,W;Z',W')$ of the two-dimensional Klein–Gordon operator as
\begin{eqnarray}
&&\Psi^{(1)}(Z,W)=-\int\!\!\int_{Z'>|W'|,Z'\geq0}dZ'dW'G(Z,W;Z',W')\frac{1}{Z'^2-W'^2}\nonumber\\
&&\qquad\qquad\qquad\qquad\qquad\qquad\times\left(\frac{12\gamma}{r_s^4}e^{7X'-5Y'}+\frac{144\eta}{r_s^6}e^{10X'-8Y'}\right)\Psi^{(0)}(Z',W'),
\end{eqnarray}
where the Green’s function is given by
\begin{align}
\label{Green function}
G(Z,W;Z',W')&=\int \frac{d^2k}{\pi^2}\frac{e^{-i\left(k_2(W-W')-k_1(Z-Z')\right)}}{k_2^2-k_1^2-4},\\
&=-\theta(s)H_0^{(2)}(2\sqrt{s})-\frac{2i}{\pi}\theta(-s)K_0(2\sqrt{-s}),
\end{align}
where $H_0^{(2)}(x)$ is the Hankel function of second kind, $K_0(x)$ is the modified Bessel function and $s=(W-W')^2-(Z-Z')^2$ for the interval \cite{DiSessa:1974xd,Hong_Hao_2010}. The Green function satisfies the equation
\begin{equation}
\left(\frac{1}{4}\frac{\partial^2}{\partial Z^2}-\frac{1}{4}\frac{\partial^2}{\partial W^2}-1\right)G(Z,W;Z',W')=\delta(Z-Z')\delta(W-W').
\end{equation}

After inserting the explicit form of $\Psi^{(0)}$ and changing variables, this integral becomes
\begin{align}
\label{wave function 1}
\Psi^{(1)}(Z,W)&=-24\pi r_s\int\!\!\int_{Z^\prime>|W^\prime|} \frac{dZ'dW'}{Z'^2-W'^2}G(Z,W;Z',W')\left(\frac{\gamma}{r_s^4}e^{7X^\prime-5Y^\prime}+\frac{12\eta}{r_s^6}e^{10X^\prime-8Y^\prime}\right)\nonumber\\
&\qquad\qquad\qquad\qquad\times e^{Y^\prime}\sinh X^\prime e^{-2r_se^{Y\prime}\cosh X^\prime}\nonumber\\
&=-24\pi r_s\int\!\!\int_{Z^\prime>|W^\prime|} dZ'dW'G(Z,W;Z',W')\frac{W'(Z'+W')^7}{(Z'^2-W'^2)^7}\left(\gamma+12\frac{\eta r_s(Z'+W')^3}{(Z'^2-W'^2)^3}\right)\nonumber\\
&\qquad\qquad\qquad\qquad\times e^{-2Z^\prime}.
\end{align}
The solution to the Wheeler–DeWitt (WDW) equation is highly sensitive to the choice of boundary conditions. In this work, we adopt the Feynman boundary condition as the physically appropriate prescription. The perturbative construction of the WDW solution satisfying this condition was discussed in Ref.~\cite{deCesare:2015vca}.

In Yeom’s annihilation-to-nothing scenario, the resolution of the singularity is not inferred directly from the asymptotic location of the classical singularity itself, but is instead interpreted as arising from the interference of wave packets at $(X = 0)$. In this picture, wave packets propagating along opposite arrows of time collide at $X = 0$ and undergo wave-packet annihilation process, leading to the vanishing of the Wheeler--DeWitt wave function.

Therefore, the analysis of the wave function in the vicinity of $X = 0$ is not an arbitrary choice but plays an essential role in assessing the validity of Yeom’s mechanism. The primary objective of this work is to examine whether this annihilation process persists once corrections from effective field theory (EFT) are taken into account.

In general, singularity resolution is expected to originate from genuinely nonperturbative quantum gravitational effects, and such behavior is unlikely to arise within the domain of a low-energy effective theory. In what follows, we analyze how the wave function behaves at $X = 0$ when higher-curvature corrections are incorporated within the EFT framework.

For clarity of presentation, we omit the detailed algebra and present the full derivation in Appendix~A. The perturbative WDW wave function in the region $X\approx0$ and $Y<<-1$ is given by
\begin{eqnarray}
\label{wave function 2}
&&\Psi^{(1)}\big|_{X=0,\,Y\ll-1}
=-\frac{16777216\sqrt{\pi}\,r_s\gamma}{e^8}
\Big(\frac{\ln\epsilon}{\epsilon^{12}}+\frac{1}{12}\frac{1}{\epsilon^{12}}+\frac{\gamma_E}{\epsilon^{12}}-\frac{\ln a}{a^{12}}-\frac{1}{12}\frac{1}{a^{12}}-\frac{\gamma_E}{a^{12}}\Big)\nonumber\\
&&\qquad\qquad\qquad\qquad +\frac{414998793616\sqrt{22\pi}\,r_s^2\eta}{e^{11}}
\Big(\frac{\ln\epsilon}{\epsilon^{18}}+\frac{1}{18}\frac{1}{\epsilon^{18}}+\frac{\gamma_E}{\epsilon^{18}}-\frac{\ln a}{a^{18}}-\frac{1}{18}\frac{1}{a^{18}}-\frac{\gamma_E}{a^{18}}\Big).\nonumber\\
&&
\end{eqnarray}
Here, $\epsilon$ denotes the lower cutoff of the $\rho$-integration, while $a$ represents the upper cutoff chosen to delimit the region that gives the dominant contribution to the integral. With these cutoffs, the leading small-$\epsilon$ behavior is dominated by $\epsilon^{-12}\ln\epsilon$ and $\epsilon^{-18}\ln\epsilon$ terms coming from the two contributions respectively.

In conventional field theory, such divergences are typically cancelled by counterterms and renormalized order by order in the loop expansion. Unlike in full quantum field theory, however, the minisuperspace model does not deal with divergences arising from loop corrections in the same way, and therefore the standard renormalization procedure cannot be applied. Within this context, it is highly nontrivial in the minisuperspace model to absorb a singularity of the form $\epsilon^{-12}$ or $\epsilon^{-18}$ into the limited set of parameters of the effective action and obtain a well-defined quantity; it is expected to be practically impossible without introducing ad hoc fine-tuning or explicitly specifying self-adjoint extensions.

Such divergences indicate a breakdown of the perturbative expansion within the low-energy EFT framework and, consequently, provide a strong argument against the robustness of the \emph{annihilation-to-nothing} scenario.

Therefore, the classical singularity is not expected to be resolved within the framework of the low-energy EFT, and its resolution is suggested to require a framework capable of appropriately describing UV physics, such as the introduction of degrees of freedom beyond those captured by general relativity or dynamics that remain consistently defined into the ultraviolet regime.

\section{Wheeler-DeWitt wave function for quantum higher curvature corrections}

In the previous section we examined classical higher-curvature corrections to the WDW wave function in the black-hole interior within the framework of effective field theory. This analysis indicates that, within the regime of validity of the low-energy effective theory, the classical singularity remains unresolved, suggesting that its resolution requires a UV-complete theory of quantum gravity.

In this section, we consider four-dimensional quadratic gravity as an effective field theory and analyze how quantum corrections modify the WDW wave function in this framework. While quadratic curvature terms are classically irrelevant in four dimensions, in the sense that they do not modify the classical equations of motion, they can nevertheless contribute at the quantum level. By treating quadratic gravity as an effective field theory, it becomes possible to systematically analyze the quantum contributions arising from quadratic curvature terms to the WDW wave function. The action of quadratic gravity is given by 
\begin{eqnarray}
S=\int d^4x\sqrt{-g}(R+\gamma R^2+\eta R_{\mu\nu}R^{\mu\nu}+\lambda E_{4}),
\end{eqnarray}
where $E_{4}$ is the topological Euler term,
\begin{eqnarray}
E_{4}=R^2-4R_{\mu\nu}R^{\mu\nu}+R_{\mu\nu\rho\sigma}R^{\mu\nu\rho\sigma}.
\end{eqnarray}
In four-dimensional spacetime, this topological term becomes a total derivative and thus does not contribute to the local equations of motion.  
The physical dynamics are therefore governed by the Ricci scalar and Ricci tensor terms.

Inside the event horizon, we consider a Kantowski--Sachs–type metric, following the same procedure as in the previous section,
\begin{eqnarray}
ds^2=g_{\mu\nu}dx^{\mu}dx^{\nu}=-e^{2\alpha(t)+4\beta(t)}dt^2+e^{2\alpha(t)}dr^2+r_s^2e^{2\beta(t)}d\Omega_2^2.
\end{eqnarray}

By substituting this metric into the quadratic gravity action, we obtain
\begin{eqnarray}
&&S=\int d^4x\Bigl[-2\dot{\alpha}\dot{\beta}-\dot{\beta}^2+r_s^2e^{2\alpha+2\beta}+\frac{\gamma}{r_s^4}(-2\dot{\alpha}\dot{\beta}-\dot{\beta}^2+r_s^2e^{2\alpha+2\beta}+\ddot{\alpha}+2\ddot{\beta})^2e^{-2(\alpha+2\beta)}\nonumber\\
&&\qquad\qquad\ \ +\frac{\eta}{r_s^4}\Bigl(\ddot{\alpha}+(4\dot{\alpha}\dot{\beta}+2\dot{\beta}^2-\ddot{\alpha}-2\ddot{\beta})^2 +2(\ddot{\beta}+r_s^2e^{2Y})^2\Bigr)e^{-2(\alpha+2\beta)}\Bigr].
\end{eqnarray}

To simplify the expression, it is convenient to introduce the field redefinitions $X(t)=\alpha(t)$ and $Y(t)=\alpha(t)+\beta(t)$, as in the previous section. 
In terms of these new variables, the action takes the form
\begin{eqnarray}
&&S=\int d^4x\Bigl[\dot{X}^2-\dot{Y}^2+r_s^2e^{2Y}+\frac{\gamma}{r_s^4} e^{2X-4Y}(\dot{X}^2-\dot{Y}^2+r_s^2e^{2Y}-\ddot{X}+2\ddot{Y})^2\nonumber\\
&&\qquad\qquad\ \ +\frac{\eta}{r_s^4}e^{2X-4Y}\Bigl(\ddot{X}^2+(-2\dot{X}^2+2\dot{Y}^2+\ddot{X}-2\ddot{Y})^2 +2(-\ddot{X}+\ddot{Y}+r_s^2e^{2Y})^2\Bigr)\Bigr].
\end{eqnarray}
 In this work, we basically employ the perturbative analysis with small parameter $\gamma,\eta=\mathcal{O}(\epsilon)$. In this Lagrangean, the motion fo equations on the zeroth order are
\begin{eqnarray}
&&\qquad\ \ \ \ \ \ddot{X}=0,\\
&&\ddot{Y}+r_s^2e^{2Y}=0,
\end{eqnarray}
which reproduce the Einstein gravity limit without higher-derivative effects.

In the framework of effective field theory, no new physical degrees of freedom, which are absent in Einstein gravity, appear at this perturbative order. To remove spurious higher-derivative terms, we perform field redefinitions as
\begin{eqnarray}
\label{X1}
&&X\rightarrow X-\frac{\gamma}{2r_s^4}e^{2X-4Y}\left(2\dot{X}^2-2\dot{Y}^2+2r_s^2e^{2Y}-\ddot{X}+3\ddot{Y}\right)\nonumber\\
&&\ \ \ \ \ \ \ -\frac{2\eta}{r_s^4}e^{2X-4Y}(\dot{X}^2-\dot{Y}^2-\ddot{X}+\ddot{Y}),\\
\label{Y1}
&&Y\rightarrow Y-\frac{2\gamma}{r_s^4}e^{2X-4Y}(\dot{X}^2-\dot{Y}^2+\ddot{Y})-\frac{\eta}{r_s^2}(4\dot{X}^2-4\dot{Y}^2-2\ddot{X}+3\ddot{Y}-r_s^2e^{2Y}).
\end{eqnarray}
These transformations eliminate all second-derivative contributions from the Lagrangian, yielding an equivalent, low-energy description in which ghost-like excitations are integrated out.

Thus, the resulting simplified action becomes
\begin{eqnarray}
&&S=\int d^4x\Bigl[\dot{X}^2-\dot{Y}^2+r_s^2e^{2Y}+\frac{\gamma}{r_s^4} e^{2X-4Y}(\dot{X}^2-\dot{Y}^2+r_s^2e^{2Y})^2-\frac{4\gamma}{r_s^2}e^{2X-2Y}(\dot{X}^2-\dot{Y}^2)\nonumber\\
&&\qquad\qquad\ \ +\frac{4\eta}{r_s^4}e^{2X-4Y}\left(\dot{X}^2-\dot{Y}^2\right)\left(\dot{X}^2-\dot{Y}^2-2r_s^2e^{2Y}\right)+4\eta e^{2X}\Bigr]\nonumber\\
&&\ \ \ =\int d^4xL(\dot{X},X,\dot{Y},Y).
\end{eqnarray}

\subsection{Hamiltonian constraint and Wheeler-DeWitt wave function}

Let us perform the Hamiltonian analysis by considering this Lagrangian, which is physically equivalent to the original one. From this action, we can compute the conjugate momenta
\begin{eqnarray}
&&\Pi_X=2\dot{X}+\frac{2\gamma}{r_s^4}e^{2X-4Y}\Pi_X\left(\frac{1}{4}\Pi_X^2-\frac{1}{4}\Pi_Y^2-r_s^2e^{2Y}\right)\nonumber\\
&&\qquad\ \ \ +\frac{8\eta}{r_s^4}e^{2X-4Y}\Pi_X\left(\frac{1}{4}\Pi_X^2-\frac{1}{4}\Pi_Y^2-r_s^2e^{2Y}\right)+\mathcal{O}(\epsilon^2),\\
&&\Pi_Y=-2\dot{Y}+\frac{2\gamma}{r_s^4}e^{2X-4Y}\Pi_Y\left(\frac{1}{4}\Pi_X^2-\frac{1}{4}\Pi_Y^2-r_s^2e^{2Y}\right)\nonumber\\
&&\qquad\ \ \ +\frac{8\eta}{r_s^4}e^{2X-4Y}\Pi_Y\left(\frac{1}{4}\Pi_X^2-\frac{1}{4}\Pi_Y^2-r_s^2e^{2Y}\right)+\mathcal{O}(\epsilon^2).
\end{eqnarray}
Then, the Hamiltonian density is 
\begin{eqnarray}
\label{Hamiltonian}
&&H=\int dx^3\Bigl[\Pi_X\dot{X}+\Pi_Y\dot{Y}-L(X,\dot{X},Y,\dot{Y})\Bigr]\nonumber\\
&&\ \ \ =\int dx^3\Bigl[\frac{1}{4}\Pi_X^2-\frac{1}{4}\Pi_Y^2-r_s^2e^{2Y}-\frac{1}{r_s^4}(\gamma+4\eta)\left(\frac{1}{4}\Pi_X^2-\frac{1}{4}\Pi_Y^2-r_s^2e^{2Y}\right)^2e^{2X-4Y}\Bigr].
\end{eqnarray}
The Hamiltonian can be written in terms of the Hamiltonian and momentum constraints, from which the WDW equation is obtained. In the canonical quantization scheme, this equation arises from promoting the Hamiltonian constraint to a quantum operator. The canonical formulation of four-dimensional quadratic gravity, including the explicit forms of these constraints, has been discussed in detail in Refs.~\cite{Kaku:1982xt,Buchbinder:1987vp,Kluson:2013hza}.

In the next step, using the Weyl ordering, we replace the classical observables with quantum operators. 
In this prescription, a classical observable $A(q,p)$ is promoted to an operator according to
\begin{eqnarray}
\hat{A}(\hat{q},\hat{\Pi})=e^{\frac{1}{2i}\frac{\partial}{\partial q}\frac{\partial}{\partial \Pi}}A_0(q,\Pi)\mid_{q=\hat{q},\Pi=\hat{\Pi}}.
\end{eqnarray}
By expandong the second term of Eq.(\ref{Hamiltonian}), we find that:
\begin{eqnarray}
&&\ \ \ \left(\frac{1}{4}\Pi_X^2-\frac{1}{4}\Pi_Y^2-r_s^2e^{2Y}\right)^2e^{2X-4Y}\nonumber\\
&&=\frac{1}{16}\Pi_X^4e^{2X-4Y}+\frac{1}{16}\Pi_Y^2e^{2X-4Y}+r_s^4e^{2X}-\frac{1}{8}\Pi_X^2\Pi_Y^2e^{2X-4Y}\nonumber\\
&&\ \ \ -\frac{1}{2}r_s^2\Pi_X^2e^{2X-2Y}+\frac{1}{2}\Pi_Y^2r_s^2e^{2X-2Y}.
\end{eqnarray}
Each term in the right-hand side is then quantized according to the Weyl ordering as follows.
\begin{eqnarray}
&&\Pi_X^4e^{2X-4Y}\rightarrow e^{2\hat{X}-4\hat{Y}}\hat{\Pi}_X^4+\frac{4}{i}e^{2\hat{X}-4\hat{Y}}\hat{\Pi}_X^3-6e^{2\hat{X}-4\hat{Y}}\hat{\Pi}_X^2\nonumber\\
&&\qquad\qquad\ \ \ \ \ \ -\frac{4}{i}e^{2\hat{X}-4\hat{Y}}\hat{\Pi}_X+e^{2\hat{X}-4\hat{Y}},\\
&&\Pi_Y^4e^{2X-4Y}\rightarrow e^{2\hat{X}-4\hat{Y}}\hat{\Pi}_Y^4-\frac{8}{i}e^{2\hat{X}-4\hat{Y}}\hat{\Pi}_Y^3-24e^{2\hat{X}-4\hat{Y}}\hat{\Pi}_Y^2\nonumber\\
&&\qquad\qquad\ \ \ \ \ \ +\frac{32}{i}e^{2\hat{X}-4\hat{Y}}\hat{\Pi}_Y+16e^{2\hat{X}-4\hat{Y}},\\
&&\Pi_X^2\Pi_Y^2e^{2X-4Y}\rightarrow e^{2\hat{X}-4\hat{Y}}\hat{\Pi}_X^2\hat{\Pi}_Y^2-\frac{4}{i}e^{2\hat{X}-4\hat{Y}}\hat{\Pi}_X^2\hat{\Pi}_Y-4e^{2\hat{X}-4\hat{Y}}\hat{\Pi}_X^2\nonumber\\
&&\qquad\qquad\ \ \ \ \ \ +\frac{2}{i}e^{2\hat{X}-4\hat{Y}}\hat{\Pi}_X\hat{\Pi}_Y^2+8e^{2\hat{X}-4\hat{Y}}\hat{\Pi}_X\hat{\Pi}_Y-\frac{8}{i}e^{2\hat{X}-4\hat{Y}}\hat{\Pi}_X\nonumber\\
&&\qquad\qquad\ \ \ \ \ \ -e^{2\hat{X}-4\hat{Y}}\hat{\Pi}_Y^2+\frac{4}{i}e^{2\hat{X}-4\hat{Y}}\hat{\Pi}_Y+4e^{2\hat{X}-4\hat{Y}}\\
&&\Pi_X^2e^{2X-2Y}\rightarrow e^{2\hat{X}-2\hat{Y}}\hat{\Pi}_X^2+\frac{2}{i}e^{2\hat{X}-2\hat{Y}}\hat{\Pi}_X-e^{2\hat{X}-2\hat{Y}},\\
&&\Pi_Y^2e^{2X-2Y}\rightarrow e^{2\hat{X}-2\hat{Y}}\hat{\Pi}_Y^2-\frac{2}{i}e^{2\hat{X}-2\hat{Y}}\hat{\Pi}_Y-e^{2\hat{X}-2\hat{Y}}.
\end{eqnarray}
Because the spacetime under consideration is static, the total Hamiltonian becomes proportional to the Hamiltonian constraint. Then, the quantum Hamiltonian constraint $\hat{\mathcal{H}}$ can be written in terms of quantum operators as
\begin{eqnarray}
&&\hat{\mathcal{H}}=\left(\frac{1}{4}\frac{\partial^2}{\partial X^2}-\frac{1}{4}\frac{\partial^2}{\partial Y^2}+r_s^2e^{2\hat{Y}}\right) +(\gamma+4\eta)e^{2\hat{X}}\nonumber\\
&&\qquad\ +\frac{1}{4r_s^2}(\gamma+4\eta)e^{2\hat{X}-2\hat{Y}}\Bigl(\frac{\partial^2}{\partial X^2}+\frac{\partial}{\partial X}+1\Bigr)\nonumber\\
&&\qquad\ +\frac{1}{16r_s^4}(\gamma+4\eta)e^{2\hat{X}-4\hat{Y}}\Bigl(\frac{\partial^4}{\partial X^4}+4\frac{\partial^3}{\partial X^3}+6\frac{\partial^2}{\partial X^2}+4\frac{\partial}{\partial X}+1\Bigr)\nonumber\\
&&\qquad\ -\frac{1}{16r_s^4}(\gamma+4\eta)e^{2\hat{X}-4\hat{Y}}\Bigl(\frac{\partial^4}{\partial X^2\partial Y^2}-4\frac{\partial^3}{\partial X^2\partial Y}+\frac{\partial^2}{\partial X^2}+2\frac{\partial^3}{\partial X\partial Y^2}\nonumber\\
&&\qquad\ -8\frac{\partial^2}{\partial X\partial Y}+8\frac{\partial}{\partial X}+\frac{\partial^2}{\partial Y^2}-4\frac{\partial}{\partial Y}+4\Bigr)\nonumber\\
&&\qquad\ -\frac{1}{4r_s^2}(\gamma+4\eta)e^{2\hat{X}-2\hat{Y}}\Bigl(\frac{\partial^2}{\partial Y^2}-\frac{\partial}{\partial Y}+1\Bigr)\nonumber\\
&&\qquad\ +\frac{1}{16r_s^4}(\gamma+4\eta)e^{2\hat{X}-4\hat{Y}}\Bigl(\frac{\partial^4}{\partial Y^4}-8\frac{\partial^3}{\partial Y^3}+24\frac{\partial^2}{\partial Y^2}-32\frac{\partial}{\partial Y}+16\Bigr),
\end{eqnarray}
where $\hat{\Pi}_X=-i\frac{\partial}{\partial X}$ and $\hat{\Pi}_Y=-i\frac{\partial}{\partial Y}$.\\

Now, the Hamiltonian constraint derives the perturbative WDW equation for the wave function $\Psi(X,Y)$,
\begin{eqnarray}
\hat{\mathcal{H}}^{(0)}\Psi^{(1)}(X,Y)=-\hat{\mathcal{H}}^{(1)}\Psi^{(0)}(X,Y),
\end{eqnarray}
where the zeroth- and first-order contributions to the Hamiltonian operator are denoted by $\hat{\mathcal{H}}^{(0)}$ and $\hat{\mathcal{H}}^{(1)}$, respectively. Here, $\hat{\mathcal{H}}^{(0)}$ represents the leading-order Hamiltonian, corresponding to Einstein gravity, while $\hat{\mathcal{H}}^{(1)}$ encodes the quadratic-curvature corrections arising from the higher-derivative terms.
\begin{eqnarray}
&&\hat{\mathcal{H}}^{(0)}=\left(\frac{1}{4}\frac{\partial^2}{\partial X^2}-\frac{1}{4}\frac{\partial^2}{\partial Y^2}+r_s^2e^{2\hat{Y}}\right),\\
&&\hat{\mathcal{H}}^{(1)}= (\gamma+4\eta)e^{2\hat{Y}}+\frac{1}{4r_s^2}(\gamma+4\eta)e^{2\hat{X}-2\hat{Y}}\Bigl(\frac{\partial^2}{\partial X^2}+\frac{\partial}{\partial X}+1\Bigr)\nonumber\\
&&\qquad\ \ \ \ +\frac{1}{16r_s^4}(\gamma+4\eta)e^{2\hat{X}-4\hat{Y}}\Bigl(\frac{\partial^4}{\partial X^4}+4\frac{\partial^3}{\partial X^3}+6\frac{\partial^2}{\partial X^2}+4\frac{\partial}{\partial X}+1\Bigr)\nonumber\\
&&\qquad\ \ \ \ -\frac{1}{16r_s^4}(\gamma+4\eta)e^{2\hat{X}-4\hat{Y}}\Bigl(\frac{\partial^4}{\partial X^2\partial Y^2}-4\frac{\partial^3}{\partial X^2\partial Y}+\frac{\partial^2}{\partial X^2}+2\frac{\partial^3}{\partial X\partial Y^2}\nonumber\\
&&\qquad\ \ \ \ -8\frac{\partial^2}{\partial X\partial Y}+8\frac{\partial}{\partial X}+\frac{\partial^2}{\partial Y^2}-4\frac{\partial}{\partial Y}+4\Bigr)\nonumber\\
&&\qquad\ \ \ \ -\frac{1}{4r_s^2}(\gamma+4\eta)e^{2\hat{X}-2\hat{Y}}\Bigl(\frac{\partial^2}{\partial Y^2}-\frac{\partial}{\partial Y}+1\Bigr)\nonumber\\
&&\qquad\ \ \ \ +\frac{1}{16r_s^4}(\gamma+4\eta)e^{2\hat{X}-4\hat{Y}}\Bigl(\frac{\partial^4}{\partial Y^4}-8\frac{\partial^3}{\partial Y^3}+24\frac{\partial^2}{\partial Y^2}-32\frac{\partial}{\partial Y}+16\Bigr).
\end{eqnarray}
As in the previous section, the first-order correction to the wave function, $\Psi^{(1)}(Z,W)$, can be expressed formally in terms of the Green’s function (\ref{Green function}) as follows:
\begin{eqnarray}
&&\Psi^{(1)}(Z,W)=-\int\!\!\int_{Z^\prime>|W^\prime|,\ Z'\geq0}dZ'dW'G(Z,W;Z',W')\frac{1}{Z'^2-W'^2}\hat{H}^{(1)}\Psi^{(0)}(Z',W')\nonumber\\
&&\qquad\qquad\ \ \ =\int\!\!\int_{Z^\prime>|W^\prime|,\ Z'\geq0}dZ'dW'G(Z,W;Z',W')\frac{1}{Z'^2-W'^2}J(Z',W'),
\end{eqnarray}
where $J(Z,W)$ is given by
\begin{align}
J(Z,W)&\equiv-\frac{(\gamma+4\eta)\pi}{8(Z-W)^3(Z+W)}e^{-2Z}\Bigl(-28W^4+16W^5-29W^3(-2+Z^2)\nonumber\\
&\ \ \ -2Z(-3+14Z+2Z^2)+Z^2W(-29+14Z^2)+W^2(-49+24Z+20Z^2)\Bigr).
\end{align}
The intermediate steps of the calculation are omitted here for clarity and are summarized in Appendix~B. We explicitly evaluate the first-order correction to the WDW wave function in the vicinity of $X=0$ and $Y<<-1$. The expression becomes
\begin{eqnarray}
\label{wave function}
&&\Psi^{(1)}|_{X=0,Y<<-1}\sim -i\frac{(\gamma+4\eta)\sqrt{2\pi}}{24e^3}\left(27\sqrt{3}-\frac{2464}{e}+\frac{30625\sqrt{5}}{8e^2}-\frac{3888\sqrt{6}}{e^3}+\frac{117649\sqrt{7}}{32e^4}\right)\nonumber\\
&&\qquad\qquad\qquad\ \ \ \ \ \ \times\left(\frac{\ln\epsilon}{\epsilon^{6}}+\frac{1}{6}\frac{1}{\epsilon^6}+\frac{\gamma_E}{\epsilon^6}-\frac{\ln a}{a^{6}}-\frac{1}{6}\frac{1}{a^6}-\frac{\gamma_E}{a^6}\right),
\end{eqnarray}
where $\epsilon$ denotes the lower cutoff of the $\rho$-integration, while $a$ represents the upper cutoff chosen to delimit the region that gives the dominant contribution to the integral as the previous section.

The analysis presented here yields the same conclusion, providing further support for the breakdown of the low-energy EFT description near the singularity. We therefore suggest that the resolution of the classical singularity requires a UV-complete theory of quantum gravity. 

\section{Summary and Discussion}

In this paper, we have reexamined the Wheeler–DeWitt (WDW) wave function in the interior region of black hole spacetimes from the perspective of effective field theory (EFT). Our aim was to show that the classical singularity cannot be resolved within the EFT framework and that any genuine resolution inevitably requires a UV-complete theory of quantum gravity.

To support this viewpoint, we incorporated higher-curvature corrections into the Einstein–Hilbert action and derived the corresponding modified WDW equation in the minisuperspace approximation. These curvature-squared and curvature-cubed terms deform the potential in the Hamiltonian constraint and lead to divergent contributions to the WDW wave function. As a result, it is suggested that the annihilation behavior in which the WDW wave function vanishes at $X=0$ is a property specific to the WDW quantization based on general relativity without higher-curvature corrections. By contrast, once EFT corrections are incorporated, this behavior is no longer maintained, and divergences appear in the WDW wave function, indicating a breakdown of the effective field theory description itself.

The key outcome of our analysis is that, by consistently including higher-curvature corrections in an EFT extension, the WDW wave function develops divergences, revealing a breakdown of the effective field theory description itself. This suggests that resolving the singularity requires either new physical degrees of freedom beyond those captured by general relativity or dynamics that are consistently defined up to arbitrarily high energy scales.

The broader implication is that general relativity, when regarded as a low-energy EFT, is not capable of resolving singularities. Any consistent resolution must rely on additional degrees of freedom associated with UV physics or dynamics that are consistently defined up to arbitrarily high energy scales. The deformation of the WDW dynamics highlights the limited domain of validity of the semiclassical approximation and shows that singularity resolution must ultimately be addressed within a UV-complete theory. Nevertheless, our results do not invalidate Yeom’s underlying intuition; rather, they suggest that any annihilation-to-nothing–type mechanism must be formulated within a UV-complete framework beyond the reach of EFT.

Future directions include studying nonperturbative approaches that serve as candidates for UV-complete quantum gravity, such as loop quantum gravity \cite{Modesto:2004wm,Ashtekar:2005qt,Corichi:2015xia,Sartini:2020ycs,Geiller:2020xze,Ongole:2022rqi} or asymptotic safety \cite{Weinberg:1980gg,Reuter:1996cp,Souma:1999at}. Another promising direction is to investigate spinorial or supersymmetric extensions of the WDW equation \cite{Kan:2021yoh,Kan:2021fmw,Kan:2022ism,Lopez-Dominguez:2009ept,Lopez-Dominguez:2011tcj}, where the inner-product structure may be better controlled. Such developments may clarify whether the annihilation-to-nothing scenario can genuinely contribute to singularity resolution or whether it remains a feature emerging solely within the framework of general relativity.

\appendix
\section{Derivation of the first-order perturbative Wheeler-DeWitt wave function with the classical higher curvature corrections}

In this appendix, we present the intermediate steps leading from Eq. (\ref{wave function 1}) to Eq. (\ref{wave function 2}). The derivation is straightforward but somewhat lengthy, so we collect the detailed algebra here for completeness.

Since evaluating the full wave function analytically at a general point is intractable, we instead focus on the regime $Y\ll -1$. When $X=0$ and $Y\ll -1$, one finds $Z\ll r_s$ and $W=0$. Therefore, $\Psi^{(0)}|_{X=0,Y\ll -1}$ approximately corresponds to
\begin{eqnarray}
&&\Psi^{(1)}(Z,W)\sim48ir_s\int\!\!\int_{Z^\prime>|W^\prime|} dZ'dW'K_0(2\sqrt{-s})\frac{W'(Z'+W')^7}{(Z'^2-W'^2)^7}\left(\gamma+12\frac{\eta r_s(Z'+W')^3}{(Z'^2-W'^2)^3}\right)e^{-2Z^\prime}\nonumber\\
&&\qquad\qquad\ \ \ \ \sim48ir_s\int_{\epsilon}^ad\rho\int_0^\infty d\kappa K_0(2\sqrt{-s})\frac{\sinh\kappa(\cosh\kappa+\sinh\kappa)^7}{\rho^5}\nonumber\\
&&\qquad\qquad\qquad\qquad\qquad\qquad\qquad\times\left(\gamma+12\frac{\eta r_s(\cosh\kappa+\sinh\kappa)^3}{\rho^3}\right)e^{-2\rho\cosh\kappa},
\end{eqnarray}
where we perform the change of variables
\begin{eqnarray}
Z=\rho\cosh\kappa,\ \ \ W=\rho\sinh\kappa.
\end{eqnarray}
The corresponding coordinate is $\rho\in(0,\infty),\kappa\in(-\infty,\infty)$.

To regulate the short-distance behavior of the $\rho$-integral, we introduce a cutoff at $\rho=\epsilon$ with $\epsilon \ll 1$. We then consider the hierarchy $e^{Y}\ll\epsilon$, ensuring that the modified Bessel function $K_0(2\sqrt{-s})$ is only weakly dependent on $Y$ in the integration domain. Since the integrand becomes negligible for large $\rho$, that region does not contribute to the value of the WDW wave function. We therefore introduce an $\mathcal{O}(1)$ upper cutoff $\rho=a$, which effectively restricts the integration to the region contributing significantly to the amplitude without affecting the asymptotic behavior. This allows us to focus on the dominant contribution arising from $\epsilon<\rho<a$. Moreover, because the exponential convergence ensures that the integrand is negligible for negative $\kappa$, we may consider only the contribution from the positive $\kappa$ region.

In this regime, we expand the Bessel function $K_0(2\sqrt{-s})$ for small $\rho\ (e^{Y}\ll\rho\ll1)$ as
\begin{eqnarray}
K_0(2\sqrt{-s})\sim K_0(2\rho)\sim -\ln(\rho)-\gamma_E,
\end{eqnarray}
where $\gamma_E$ is the Euler--Mascheroni constant. Substituting this expansion into the integrand yields
\begin{align}
\Psi^{(1)}(Z,W)&\sim48ir_s\int_{\epsilon}^{a}d\rho\int_0^{\infty}d\kappa
\frac{\sinh\kappa(\cosh\kappa+\sinh\kappa)^7}{\rho^5}\left(\gamma+12\frac{\eta r_s(\cosh\kappa+\sinh\kappa)^3}{\rho^3}\right)e^{-2\rho\cosh\kappa}\nonumber\\
&\qquad\qquad\qquad\qquad\times[-\ln(\rho\cosh\kappa)-\gamma_E].
\end{align}

Since the exponential factor $e^{-\rho\cosh\kappa}$ suppresses the large-$\kappa$ region, the integral is dominated by small $\rho$ and moderate $\kappa$ values. To analyze this dominant contribution more systematically, we apply the Laplace (saddle-point) approximation and use the large-$\kappa$ asymptotics.
\begin{equation}
\cosh\kappa+\sinh\kappa=e^{\kappa},\qquad
\sinh\kappa\sim\frac{1}{2}e^{\kappa},\qquad
\cosh\kappa\sim\frac{1}{2}e^{\kappa}.
\end{equation}
Applying these to the first term gives
\begin{equation}
\sinh\kappa\big(\cosh\kappa+\sinh\kappa\big)^7 e^{-2\rho\cosh\kappa}
\sim \frac{1}{2}e^{8\kappa} e^{-\rho e^{\kappa}}.
\end{equation}
It is convenient to define the phase function
\begin{equation}
\Phi(\rho,\kappa)\equiv 8\kappa-\rho e^{\kappa},
\end{equation}
and to assume $\kappa$ is large; this assumption is consistent provided
\begin{equation}
\rho \ll \frac{8}{e},
\end{equation}
so that the saddle occurs at large $\kappa$.

The saddle point $\kappa_*$ with respect to $\kappa$ is determined by $\partial_\kappa\Phi:=\Phi'=0$:
\begin{eqnarray}
\Phi'(\rho,\kappa_*)&=&8-\rho e^{\kappa_*}=0,\nonumber\\
&&\Rightarrow\quad e^{\kappa_*}=\frac{8}{\rho},\qquad
\kappa_*=\ln\!\left(\frac{8}{\rho}\right),
\end{eqnarray}
and the second derivative at the saddle is
\begin{equation}
\Phi''(\rho,\kappa_*)=-\rho e^{\kappa_*}=-8.
\end{equation}

More generally, we consider integrals of the form
\begin{equation}
I_m(\rho)=\int_{-\infty}^{\infty} d\kappa\,\rho^{p_m} c_m\,e^{\,m\kappa-\rho e^{\kappa}}.
\end{equation}
The saddle equation for this integrand is $m-\rho e^{\kappa}=0$, so
\begin{equation}
e^{\kappa_m}=\frac{m}{\rho},\qquad \kappa_m=\ln\!\left(\frac{m}{\rho}\right),
\qquad \Phi''\big|_{\kappa_m}=-\rho e^{\kappa_m}=-m.
\end{equation}
By the standard Laplace formula (Gaussian approximation around the saddle) we obtain
\begin{equation}
I_m(\rho)\approx c_m\,\rho^{p_m}\,e^{\,m\kappa_m-\rho e^{\kappa_m}}\sqrt{\frac{2\pi}{m}}
= c_m\,\rho^{p_m}\left(\frac{m}{\rho}\right)^m e^{-m}\sqrt{\frac{2\pi}{m}}.
\end{equation}

Using this result with the appropriate values of $m$ and coefficients (in particular $m=8$ and $m=11$ for the two dominant contributions), the first-order correction $\Psi^{(1)}$ evaluated at $X=0$ and $Y\ll -1$ becomes 
\begin{eqnarray}
&&\Psi^{(1)}\big|_{X=0,\,Y\ll-1}
\approx24\sqrt{2\pi}\,r_s\int_{\epsilon}^{a} d\rho\,
\frac{1}{e^8\rho^{13}}\big(-\ln\rho-\gamma_E\big)\nonumber\\
&&\qquad\qquad\qquad\qquad\times
\Big(4194304\sqrt{2}\,\gamma+311249095212\sqrt{11}\,\frac{\eta r_s}{e^3\rho^6}\Big),
\end{eqnarray}
where in the second line we have used the small-$\rho$ expansion
\begin{equation}
K_0(2\rho)\simeq -\ln\rho-\gamma_E,
\end{equation}
keeping the leading logarithmic piece and Euler's constant $\gamma_E$.

The $\rho$-integrals that appear are of the type
\begin{equation}
\int_{\epsilon}^{a}\rho^{-p}\ln\rho\,d\rho
= -\frac{\ln a}{p-1} a^{-p+1} -\frac{a^{-p+1}}{(p-1)^2}
+ \frac{\ln\epsilon}{p-1}\epsilon^{-p+1} + \frac{\epsilon^{-p+1}}{(p-1)^2},
\end{equation}
which shows explicitly the power-law divergences as $\epsilon\to0$. Carrying out these integrals for the two terms in the integrand yields
\begin{eqnarray}
&&\Psi^{(1)}\big|_{X=0,\,Y\ll-1}
=-\frac{16777216\sqrt{\pi}\,r_s\gamma}{e^8}
\Big(\frac{\ln\epsilon}{\epsilon^{12}}+\frac{1}{12}\frac{1}{\epsilon^{12}}+\frac{\gamma_E}{\epsilon^{12}}-\frac{\ln a}{a^{12}}-\frac{1}{12}\frac{1}{a^{12}}-\frac{\gamma_E}{a^{12}}\Big)\nonumber\\
&&\qquad\qquad\qquad\qquad +\frac{414998793616\sqrt{22\pi}\,r_s^2\eta}{e^{11}}
\Big(\frac{\ln\epsilon}{\epsilon^{18}}+\frac{1}{18}\frac{1}{\epsilon^{18}}+\frac{\gamma_E}{\epsilon^{18}}-\frac{\ln a}{a^{18}}-\frac{1}{18}\frac{1}{a^{18}}-\frac{\gamma_E}{a^{18}}\Big),\nonumber\\
&&
\end{eqnarray}
so that the leading small-$\epsilon$ behavior is dominated by $\epsilon^{-12}\ln\epsilon$ and $\epsilon^{-18}\ln\epsilon$ terms coming from the two contributions respectively.

The divergences that arise in the EFT formulated on minisuperspace are unlikely to be consistently renormalized within the limited parameter space of the effective action. This suggests a breakdown of the perturbative expansion itself. These results raise questions about the robustness of the annihilation-to-nothing scenario. Consequently, the classical singularity is not resolved within the framework of a low-energy effective theory, and its resolution is suggested to require a complete theory of quantum gravity that consistently incorporates ultraviolet physics.

\section{Derivation of the first-order perturbative Wheeler-DeWitt wave function with the quantum higher curvature corrections}

This appendix summarizes the detailed expansion used in Sec.~4 to obtain Eq. (\ref{wave function}). The computation involves several intermediate manipulations, which we display here for clarity.

By introducing polar-type variables $(\rho,\kappa)$, we rewrite the source $J(Z,W)$ as $\tilde{J}(\rho,\kappa)$, which is more convenient for the subsequent analysis in the asymptotic regime $\kappa\rightarrow\infty$.
\begin{align}
\hat{J}(\rho,\kappa)&=-\frac{(\gamma+4\eta)\pi}{8\rho^3(\cosh\kappa-\sinh\kappa)^2}e^{-2\rho\cosh\kappa}(-28\rho^3\sinh^4\kappa+16\rho^4\sinh^5\kappa\nonumber\\
&\ \ \ -29\rho^2\sinh^3\kappa(-2+\rho^2\cosh^2\kappa)-2\cosh\kappa(-3+14\rho\cosh\kappa+2\rho^2\cosh^2\kappa)\nonumber\\
&\ \ \ +\rho^2\cosh^2\kappa\sinh\kappa(-29+14\rho^2\cosh^2\kappa)\nonumber\\
&\ \ \ +\rho\sinh^2\kappa(-49+24\rho\cosh\kappa+20\rho^2\cosh^2\kappa)),\\
&\sim-\frac{(\gamma+4\eta)\pi}{8\rho^3}e^{2\kappa-\rho e^{\kappa}}\left(3e^{\kappa}-\frac{77}{4}\rho e^{2\kappa}+\frac{49}{8}\rho^2 e^{3\kappa}-\frac{1}{2}\rho^3 e^{4\kappa}+\frac{1}{32}\rho^4e^{5\kappa}\right).
\end{align}
Then,
\begin{align}
\Psi^{(1)}(Z,W)&\sim i\int_\epsilon^ad\rho\int_0^{\infty}d\kappa K(2\sqrt{s})\frac{(\gamma+4\eta)}{4\rho^4}e^{2\kappa-\rho e^{\kappa}}\nonumber\\
&\qquad\qquad\times\left(3e^{\kappa}-\frac{77}{4}\rho e^{2\kappa}+\frac{49}{8}\rho^2 e^{3\kappa}-\frac{1}{2}\rho^3 e^{4\kappa}+\frac{1}{32}\rho^4 e^{5\kappa}\right).
\end{align}
Using these expressions, we can explicitly evaluate the first-order correction to the wave function near the classical singularity ($X=0$, $Y\ll -1$). The source term $\hat{J}(\rho,\kappa)$ is expressed in terms of the polar-like variables, allowing a convenient expansion in powers of $\rho^{-1}$ for the asymptotic analysis.
\begin{align}
\Psi^{(1)}|_{X=0,Y<<-1}&\sim i\int_{\epsilon}^ad\rho K_0(2\rho)\frac{(\gamma+4\eta)\sqrt{2\pi}}{4e^3\rho^7}\nonumber\\
&\qquad\times\left(27\sqrt{3}-\frac{2464}{e}+\frac{30625\sqrt{5}}{8e^2}-\frac{3888\sqrt{6}}{e^3}+\frac{117649\sqrt{7}}{32e^4}\right).
\end{align}
Analogously to Appendix A, keeping only the leading logarithmic term of the modified Bessel function together with Euler’s constant, we obtain
\begin{eqnarray}
&&\Psi^{(1)}|_{X=0,Y<<-1}\sim -i\int_{\epsilon}^ad\rho\frac{(\gamma+4\eta)\sqrt{2\pi}}{4e^3\rho^7}(\ln\rho+\gamma_E)\nonumber\\
&&\qquad\qquad\qquad\qquad\qquad\times\left(27\sqrt{3}-\frac{2464}{e}+\frac{30625\sqrt{5}}{8e^2}-\frac{3888\sqrt{6}}{e^3}+\frac{117649\sqrt{7}}{32e^4}\right)\nonumber\\
&&\qquad\qquad\qquad\ \ \ \ =-i\frac{(\gamma+4\eta)\sqrt{2\pi}}{24e^3}\left(27\sqrt{3}-\frac{2464}{e}+\frac{30625\sqrt{5}}{8e^2}-\frac{3888\sqrt{6}}{e^3}+\frac{117649\sqrt{7}}{32e^4}\right)\nonumber\\
&&\qquad\qquad\qquad\ \ \ \ \ \ \times\left(\frac{\ln\epsilon}{\epsilon^{6}}+\frac{1}{6}\frac{1}{\epsilon^6}+\frac{\gamma_E}{\epsilon^6}-\frac{\ln a}{a^{6}}-\frac{1}{6}\frac{1}{a^6}-\frac{\gamma_E}{a^6}\right).
\end{eqnarray}

As in Sec.~3, the divergences arising in the minisuperspace EFT signal the breakdown of the low-energy description, implying that the classical singularity cannot be resolved without a UV-complete theory of quantum gravity.

\bibliography{references}
\bibliographystyle{JHEP.bst}

\end{document}